\documentstyle[twoside,fleqn,espcrc2,psfig]{article}

\newcommand{\AmS}{{\protect\the\textfont2
  A\kern-.1667em\lower.5ex\hbox{M}\kern-.125emS}}

\title{Experimental test for 5$^{th}$ dimension in Kaluza-Klein gravity}

\author{V. Dzhunushaliev\address{Universit\"at Potsdam, Institute
        f\"ur Mathematik, 14469, Potsdam, Germany and \\
        Dept. Theor. Phys., Kyrgyz State National University,
        Bishkek 720024, Kyrgyzstan \\
        e-mail: dzhun@rz.uni-potsdam.de and dzhun@freenet.bishkek.su}
        \thanks{VD is grateful for financial support by a Georg Forster
        Research Fellowship from the Alexander von Humboldt Foundation
        and H.-J. Schmidt for invitation to Potsdam Universit\"at for
        research.}
        and
        D. Singleton\address{Dept. of Phys. CSU Fresno, 2345 East San Ramon
        Ave.
        M/S 37 Fresno, CA 93740-8031, USA. \\
        e-mail :
        dougs@csufresno.edu}}

\begin{document}

\begin{abstract}
Several electric/magnetic charged solutions (dyons) to 
5D Kaluza-Klein gravity on the principal bundle are reviewed. 
Here we examine the possibility that these solutions
can act as quantum virtual wormholes in spacetime
foam models. By applying a sufficently large, external
electric and/or magnetic field it may be possible to ``inflate''
these solutions from a quantum to a classical state.
This effect could lead to a possible experimental
signal for higher dimensions in multidimensional gravity.
\end{abstract}

\maketitle

\section{Introduction}

An important aspect in any gravitational theory is to find
experimental tests of the theory. Even General Relativity does
not have as wide a range of experimental tests in comparision
with solid state theory, for example. In multidimensional
(MD) gravity theories this lack of experimental tests
is even more acute. Usually, tests of MD gravity theories are
connected with effects associated with the presence of
the scalar field(s) that occur in this theories.
For example, in 5D Kaluza - Klein theory it is possible to show
that variations of the 5$^{th}$ coordinate lead to changes
in the electrical charge to mass ratio of an elementary
particle. This effect is very small since no experiment has
found such a change.
\par
\begin{figure}[htb]
\vspace{9pt}
\framebox[55mm]{
\psfig{figure=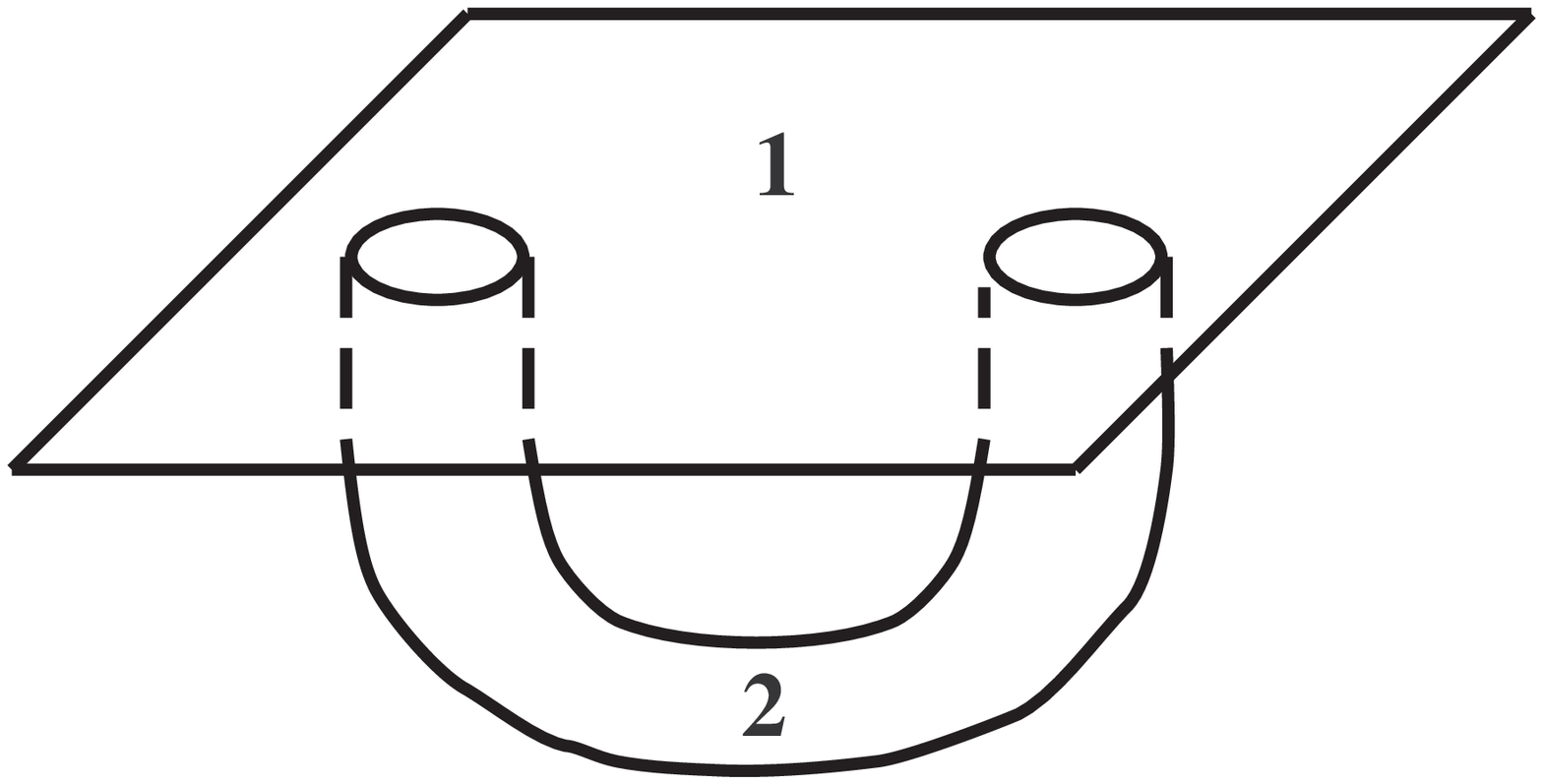,height=3cm,width=5cm}}
\caption{1 is the 5D spacetime with $G_{55}$ as a
non-dynamical variable (in this case 5D gravity is
equiavalent to the 4D gravity + electromagnetism).
2 is the 5D throat with $G_{55}$ as a dynamical
variable.}
\label{fig1}
\end{figure}

 
In this paper we propose another possible experimental test
for MD gravity. The basic idea is :
\begin{itemize}
\item
The Universe can have MD regions, see Fig.\ref{fig1}.
This means that \textit{\textbf{a piecewise compactification mechanism}} 
can exist in Nature \cite{vd1}. Piecewise compactification 
implies that some parts of the Universe
are regions where one has full MD gravity (5D in our case),
while other parts of the Universe are
ordinary 4D regions where gravity does not act
on the extra dimensions. For this mechanism to be viable it
is necessary that at the boundary between these regions
a quantum splitting off of the 5$^{th}$ dimension occurs.
In regions where gravity propagates in all the dimensions
the Universe will appear as a true 5D spacetime.
In the regions where gravity does not propagate
into the extra dimension one has ordinary
4D spacetime plus the gauge fields of the fibre.
\item
These MD regions can be inflated from the space-time foam in the
presence of very strong external electric+magnetic fields.
\end{itemize}

\section{Wormhole and flux tube solutions in 5D gravity.}

In 5D Kaluza - Klein theory there are wormholelike and
flux tube solutions \cite{vds3}.
The general form of the metric is:
\begin{eqnarray}
ds^2 & = & e^{2\nu (r)} dt^{2} - r_0^2e^{2\psi (r) - 2\nu (r)}\times
\nonumber \\
&& \left [d\chi + \omega (r)dt + \cos \theta d\varphi \right ]^2
\nonumber \\
& - & dr^{2} - a(r)(d\theta ^{2} +
\sin ^{2}\theta  d\varphi ^2),
\label{3}
\end{eqnarray}
where $\chi $ is the 5$^{th}$ coordinate;
$A_t = \omega$ and $A_{\phi}=\cos \theta$ are the 4D 
electromagnetic potentials.
\par
\begin{figure}[htb]
\vspace{9pt}
\framebox[55mm]{
\psfig{figure=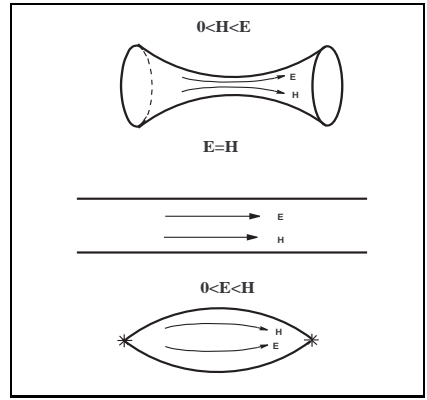,height=5cm,width=4cm}}
\caption{The different types of 5D solutions depends on the
relative strengths of the electric and magnetic fields.}
\label{fig1a}
\end{figure}

A detailed analytical and numerical investigation
of this metric gives the following
spacetime configurations, whose global structure
depends on the relationship  between the electric and
magnetic fields (see Fig.\ref{fig1a}):
\begin{enumerate}
\item
$0 \le H_{KK} < E_{KK}$. The corresponding solution
is \textbf{\textit{a WH-like object}}
located between two surfaces at $r = \pm r_0$. The cross-sectional
size of this solution (given by $a(r)$) increases as $r$ goes from $0$
to $\pm r_0$. The throat between the $r = \pm r_0$ surfaces is
filled with electric and/or magnetic flux.
\item
$H_{KK} = E_{KK}$. In this case the solution is
\textbf{\textit{an infinite flux tube}} filled
with constant electrical and magnetic fields,
with the charges at $\pm \infty$. The cross-sectional
size of this solution is constant ($ a= const.$). In Refs.
\cite{dzh1} \cite{vds3} an exact, analytical form of this
solution was found. This
solution is almost identical to the 4D Levi-Civita flux
tube solution \cite{levi} except the strength of the magnetic and
electric fields are equal, while in the Levi-Civita
solution the two fields can take on any relative value
with respect to one another. The restriction that the
electric charge equals the magnetic charge is reminiscent
of other higher dimensional soliton solutions \cite{perry}. 
The form of this infinite flux tube
configuration also has similarities to
the Anti-de Sitter (AdS) ``throat region''  that one
finds by stacking a large number of D3-branes \cite{jm}.
\item
$0 < E_{KK} \le H_{KK}$. In this case we have
\textbf{\textit{a finite flux tube}}
between two (+) and (-) magnetic and/or electric
charges, which are located at $\pm r_0$. The longitudinal
size of this flux tube is finite, but now the cross
sectional size decreases as $r \rightarrow r_0$. At
$r = \pm r_0$ this solution has real singularities which
we interpret as the locations of the magnetic and/or
and electric charges.
\end{enumerate}

\section{Creation of dyonic WHs via external electromagnetic fields}

The proposed experimental signal for the extra dimensions in
MD gravity uses these solutions in the following way:
\par
\begin{figure}[htb]
\vspace{9pt}
\framebox[55mm]{
\psfig{figure=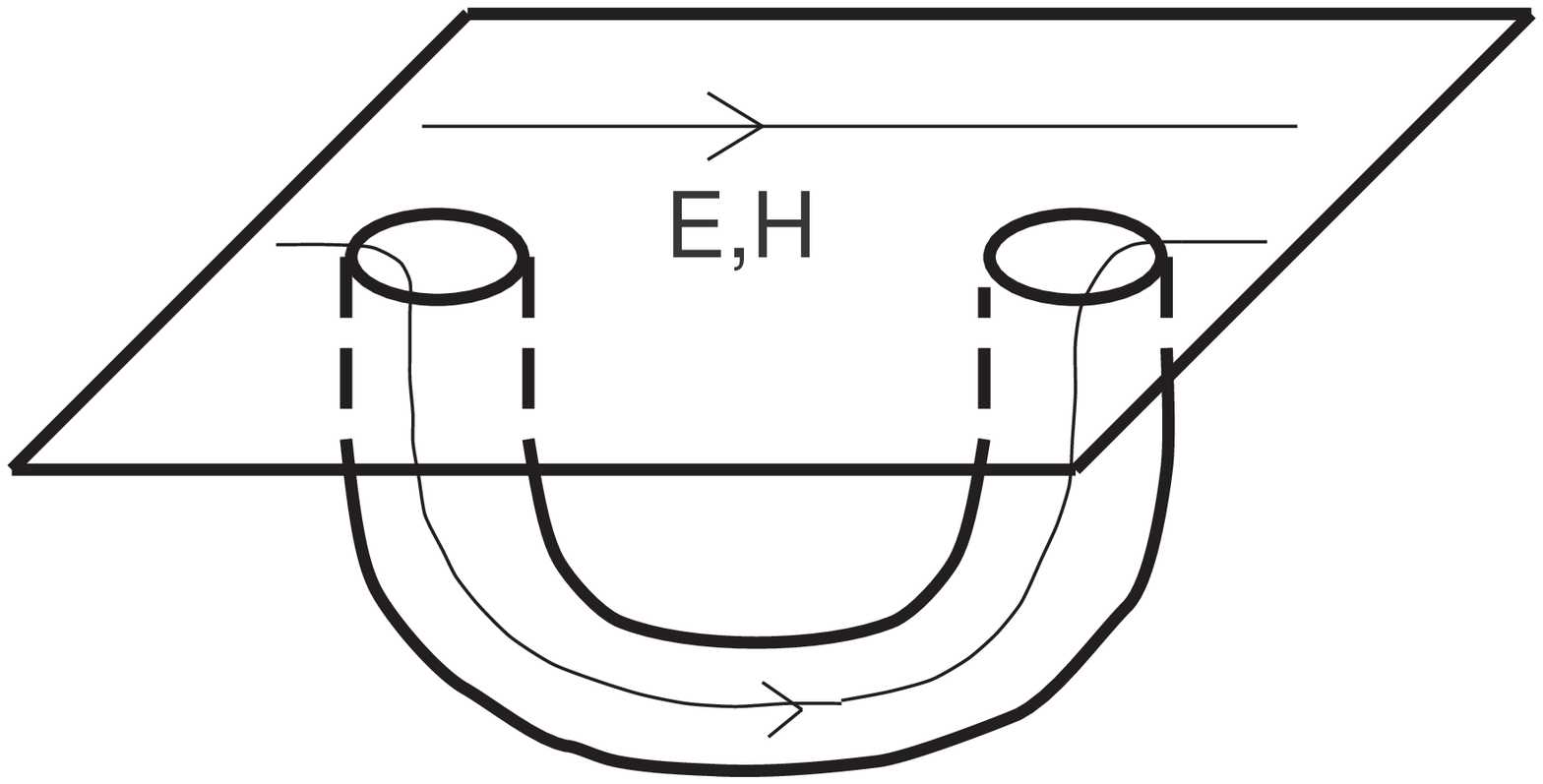,height=3cm,width=5cm}
}
\caption{The composite WH filled with the external fields.}
\label{fig2}
\end{figure}

\textit{Composite WHs can appears as a quantum handles
(quantum WHs) in the spacetime foam. These quantum
structures can be ``blown up'' or ``inflated'' from a
quantum state to a classical state by embedding it in parallel
$E$ and $H$ fields,see Fig.\ref{fig2}}
\par
These handles are taken as quantum fluctuations
in the spacetime  foam, which the externally imposed
$E$ and $H$ fields can inflate to classical states with
some probability.
\par
\begin{figure}[htb]
\vspace{9pt}
\framebox[55mm]{
\psfig{figure=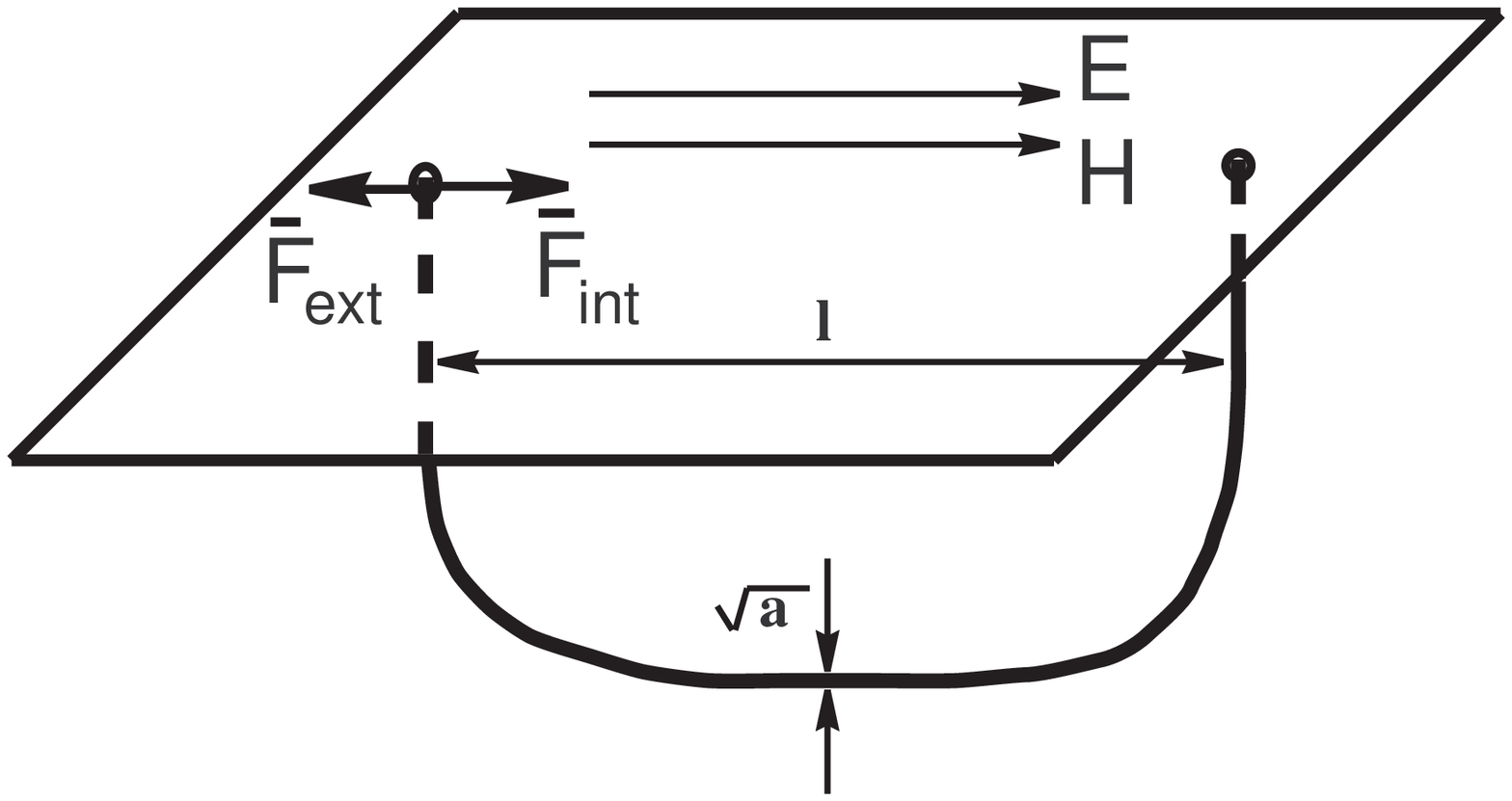,height=3cm,width=5cm}
}
\caption{The forces acting on the ends of WH.}
\label{fig3}
\end{figure}

For an external
observer these composite WHs will appear as two oppositely
charged dyons, with the charges
located on the surfaces where the 4D and 5D
spacetimes are matched, see Fig.\ref{fig3}.
Since one would like these dyonic objects to be well separated,
we will consider the case $E \approx H$ ($E>H$).
Under these conditions
the solution to Einstein's MD vacuum equations
for the metric ansatz is \cite{vds3}
\begin{eqnarray}
q \approx Q,
\label{4}\\
a \approx \frac{q^2}{2} = const,
\label{5}\\
e^{\psi} \approx e^{\nu} \approx \cosh \left( \frac{r\sqrt{2}}{q} \right),
\nonumber \\
\omega \approx \frac{\sqrt{2}}{r_0}\sinh \left( \frac{r\sqrt{2}}{q} \right)
\label{7}
\end{eqnarray}
$q$ is the electrical charge and $Q$ the magnetic charge.
Both Kaluza - Klein fields are:
\begin{equation}
E \approx H \approx \frac{q}{a} \approx \frac{2}{q}
\approx \sqrt{\frac{2}{a}}
\label{8}
\end{equation}
The cross sectional size of the WH is proportional to
$q^2$. According to this scenario the external, parallel electric
and magnetic fields should fill the virtual WH. In cgs
units the fields necessary for forming a composite WH with 
a cross sectional size $a$ are
\begin{equation}
E \approx H \approx \frac{c^2}{\sqrt G} \sqrt{\frac{2}{a}}
\label{9}
\end{equation}
In order that the charged surfaces
of the WH appear as well separated dyons
we need to require that the longitudinal
distance, $l$, between these surfaces be much larger than
the cross sectional size of the WH, $l \gg  a^ {1/2}$.
Also in order to be able to separate
the two ends of the WH as distinct electric/magnetic
charged objects one needs the external force to be much
larger than the interaction force between the
oppositely charged ends. This leads to the following condition
\begin{eqnarray}
F_{ext} &=& qE + QH \gg \frac{q^2 + Q^2}{l^2} = F_{int} \nonumber \\
&\approx& 2qE \gg \frac{2 q^2}{l^2} \rightarrow l \gg \sqrt a
\label{10}
\end{eqnarray}
If this condition holds than the oppositely charged ends will
move apart. Otherwise the ends will come back together
and annihilate back into the spacetime foam.
\par
The average value of $a$ for spacetime foam is given by the Planck size
$a^{1/2} \approx L_{Pl} \approx$ 10$^{-33}$ cm.
Thus the relevant electric field should be
$E \approx$ 3.1 $\times$ 10$^{57}$ V/cm.
This field strength is well beyond experimental capabilities 
to create. Hence one must consider quantum WHs
whose linear size satisfies $a^{1/2} \gg L_{Pl}$. The
larger $a^{1/2}$ the smaller the field strength needed.
But such large quantum WHs are most likely very rare. If $f(a)$ is the
probability density for the distribution for a
WH of cross section $a$
then $f(a)da$ gives the probability for the appearance of
quantum WH with cross section $a$. The bigger
$a$ the smaller the probability, $f(a)da$.
Also the larger the value of the external
$E$ and $H$, the smaller is the cross
sectional size $a$ of the WH that can be
inflated from the spacetime foam. Thus depending on the
unknown probability $f(a)da$ one can set up some spatial region
with parallel $E$ and $H$ fields whose magnitudes
are as large as technologically feasible, and look for
electric/magnetic charged objects whose charges are of
similar magnitude. 
\par
\section{Conclusions}
The above estimates seem to give a pessimistic view
of ``inflating'' such quantum WHs via man-made
fields. However, certain
astronomical objects are able to provide
fields which are much stronger
than those that can be made in the lab, thus increasing the
probability that a quantum WH can be inflated.
For example, neutron stars have extremely powerful
magnetic fields which are many orders of magnitude larger
than can be made in the lab. Also if the neutron star has
a companion it can capture plasma from its partner. This
captured plasma could generate localized,
but strong electric field strengths which in combination with
the powerful, global magnetic fields could lead to the ``inflation''
of the quantum WHs. To an external observer this would appear
as the creation of dyonic particles (the end of the flux tubes)
with nearly equal electric and magnetic charges. Another possibility
is that under the above scenario the magnetic and electric fields
of the compact astronomical object ({\it e.g.} a neutron star) are
not large enough for the condition in Eq. (\ref{10}) to be
satisfied. In this case the external fields might start
to inflate the quantum WH, but rather than the ends moving
apart, they might be pulled back together by the attraction
of the oppositely charged ends. The ends of the quantum WH
would thus annihilate back into the vacuum. One can show that 
just as for QCD flux tubes, that the field energy $U$ of the
present flux tubes is proportional to the length of the flux
tube, $l$. Thus, if the maximum
separation, $l$, between the ends of the flux tube were large
enough so that one had a substantial field energy, $U$,
then this annihilation could result in
an intense burst of gamma ray photons.
Such a mechanism could be linked with the gamma ray
burster phenomenon, which is often associated with
compact astronomical objects such as inspiraling, binary
neutron stars or naked singularities \cite{singh}.
\par
It is also possible that such handles were formed in the very
early Universe and then were inflated during the inflationary
phase. In order for these handles not
to have evaporated via Hawking radiation they must be
sustained by being embedded in powerful external magnetic/electric 
fields created, for example, by binary neutron stars.

\end{document}